\documentclass[preprint,aps,prd,tightenlines,superscriptaddress,nofootinbib]{revtex4}

  \usepackage{amssymb}
   \usepackage{amsmath}
    \usepackage{amsfonts}
\usepackage[dvips,final]{graphicx}
\usepackage{epsfig}
\usepackage{graphicx}
\usepackage{dcolumn}
\date{}

\date{}
\begin{document}

\newcommand{\ds}{\displaystyle}
\newcommand{\mc}{\multicolumn} 
\newcommand{\bce}{\begin{center}}
\newcommand{\ece}{\end{center}}
\newcommand{\beq}{\begin{equation}}
\newcommand{\eeq}{\end{equation}}
\newcommand{\bea}{\begin{eqnarray}}

\newcommand{\eea}{\end{eqnarray}}
\newcommand{\cont}{\nonumber\eea\bea}
\newcommand{\cl}[1]{\begin{center} {#1} \end{center}}
\newcommand{\ba}{\begin{array}}
\newcommand{\ea}{\end{array}}

\newcommand{\ab}{{\alpha\beta}}
\newcommand{\cd}{{\gamma\delta}}
\newcommand{\dc}{{\delta\gamma}}
\newcommand{\ac}{{\alpha\gamma}}
\newcommand{\bd}{{\beta\delta}}
\newcommand{\abc}{{\alpha\beta\gamma}}
\newcommand{\eps}{{\epsilon}}
\newcommand{\lam}{{\lambda}}
\newcommand{\mn}{{\mu\nu}}
\newcommand{\mpnp}{{\mu'\nu'}}
\newcommand{\Amuu}{{A_{\mu}}}
\newcommand{\Amuo}{{A^{\mu}}}
\newcommand{\Vmuu}{{V_{\mu}}}
\newcommand{\Vmuo}{{V^{\mu}}}
\newcommand{\Anuu}{{A_{\nu}}}
\newcommand{\Anuo}{{A^{\nu}}}
\newcommand{\Vnuu}{{V_{\nu}}}
\newcommand{\Vnuo}{{V^{\nu}}}
\newcommand{\Fmnu}{{F_{\mu\nu}}}
\newcommand{\Fmno}{{F^{\mu\nu}}}

\newcommand{\abcd}{{\alpha\beta\gamma\delta}}


\newcommand{\bsigma}{\mbox{\boldmath $\sigma$}}
\newcommand{\btau}{\mbox{\boldmath $\tau$}}
\newcommand{\brho}{\mbox{\boldmath $\rho$}}
\newcommand{\bpipi}{\mbox{\boldmath $\pi\pi$}} 
\newcommand{\bss}{\bsigma\!\cdot\!\bsigma}
\newcommand{\btt}{\btau\!\cdot\!\btau}
\newcommand{\bnabla}{\mbox{\boldmath $\nabla$}}
\newcommand{\bphi}{\mbox{\boldmath $\tau$}}
\newcommand{\bvarphi}{\mbox{\boldmath $\rho$}}
\newcommand{\bDelta}{\mbox{\boldmath $\Delta$}}
\newcommand{\bpsi}{\mbox{\boldmath $\psi$}}
\newcommand{\bPsi}{\mbox{\boldmath $\Psi$}}
\newcommand{\bPhi}{\mbox{\boldmath $\Phi$}}
\newcommand{\bnab}{\mbox{\boldmath $\nabla$}}
\newcommand{\bpi}{\mbox{\boldmath $\pi$}}
\newcommand{\btheta}{\mbox{\boldmath $\theta$}}
\newcommand{\bkappa}{\mbox{\boldmath $\kappa$}}
\newcommand{\bp}{\mbox{\boldmath $p$}}
\newcommand{\bq}{\mbox{\boldmath $q$}}
\newcommand{\br}{\mbox{\boldmath $r$}}
\newcommand{\bs}{\mbox{\boldmath $s$}}
\newcommand{\bk}{\mbox{\boldmath $k$}}
\newcommand{\bl}{\mbox{\boldmath $l$}}
\newcommand{\bb}{\mbox{\boldmath $b$}}
\newcommand{\bP}{\mbox{\boldmath $P$}}



\newcommand{\bT}{{\bf T}}
\newcommand{\fph}{${\cal F}$}
\newcommand{\aph}{${\cal A}$}
\newcommand{\dph}{${\cal D}$}
\newcommand{\fpi}{f_\pi}
\newcommand{\mpi}{m_\pi}
\newcommand{\Tr}{{\mbox{\rm Tr}}}
\def\Qb{\overline{Q}}
\newcommand{\delu}{\partial_{\mu}}
\newcommand{\delo}{\partial^{\mu}}
%
%
\newcommand{\up}{\!\uparrow}
\newcommand{\upup}{\uparrow\uparrow}
\newcommand{\updo}{\uparrow\downarrow}
\newcommand{\uu}{$\uparrow\uparrow$}
\newcommand{\ud}{$\uparrow\downarrow$}
\newcommand{\auu}{$a^{\uparrow\uparrow}$}
\newcommand{\aud}{$a^{\uparrow\downarrow}$}
\newcommand{\pu}{p\!\uparrow}

\newcommand{\qp}{quasiparticle}
\newcommand{\sa}{scattering amplitude}
\newcommand{\ph}{particle-hole}
\newcommand{\qcd}{{\it QCD}}
\newcommand{\integ}{\int\!d}
\newcommand{\ie}{{\sl i.e.~}}
\newcommand{\etal}{{\sl et al.~}}
\newcommand{\etc}{{\sl etc.~}}
\newcommand{\rhs}{{\sl rhs~}}
\newcommand{\lhs}{{\sl lhs~}}
\newcommand{\eg}{{\sl e.g.~}}
\newcommand{\ef}{\epsilon_F}
\newcommand{\sigt}{d^2\sigma/d\Omega dE}
\newcommand{\sige}{{d^2\sigma\over d\Omega dE}}
\newcommand{\rpaeq}{\beq
\left ( \begin{array}{cc}
A&B\\
-B^*&-A^*\end{array}\right )
\left ( \begin{array}{c}
X^{(\kappa})\\Y^{(\kappa)}\end{array}\right )=E_\kappa
\left ( \begin{array}{c}
X^{(\kappa})\\Y^{(\kappa)}\end{array}\right )
\eeq}
\newcommand{\ket}[1]{| {#1} \rangle}
\newcommand{\bra}[1]{\langle {#1} |}
\newcommand{\ave}[1]{\langle {#1} \rangle}
\newcommand{\half}{{1\over 2}}

\newcommand{\singlespace}{
    \renewcommand{\baselinestretch}{1}\large\normalsize}
\newcommand{\doublespace}{
    \renewcommand{\baselinestretch}{1.6}\large\normalsize}
\newcommand{\bftau}{\mbox{\boldmath $\tau$}}
\newcommand{\bfalpha}{\mbox{\boldmath $\alpha$}}
\newcommand{\bfgamma}{\mbox{\boldmath $\gamma$}}
\newcommand{\bfxi}{\mbox{\boldmath $\xi$}}
\newcommand{\bfbeta}{\mbox{\boldmath $\beta$}}
\newcommand{\bfeta}{\mbox{\boldmath $\eta$}}
\newcommand{\bfpi}{\mbox{\boldmath $\pi$}}
\newcommand{\bfphi}{\mbox{\boldmath $\phi$}}
\newcommand{\bfR}{\mbox{\boldmath ${\cal R}$}}
\newcommand{\bfL}{\mbox{\boldmath ${\cal L}$}}
\newcommand{\bfM}{\mbox{\boldmath ${\cal M}$}}
\def\dblint{\mathop{\rlap{\hbox{$\displaystyle\!\int\!\!\!\!\!\int$}}
    \hbox{$\bigcirc$}}}
\def\ut#1{$\underline{\smash{\vphantom{y}\hbox{#1}}}$}

\def\UNITY{{\bf 1\! |}}
\def\Pom{{\bf I\!P}}
\def\lsim{\mathrel{\rlap{\lower4pt\hbox{\hskip1pt$\sim$}}
    \raise1pt\hbox{$<$}}}         
\def\gsim{\mathrel{\rlap{\lower4pt\hbox{\hskip1pt$\sim$}}
    \raise1pt\hbox{$>$}}}         
\def\beq{\begin{equation}}
\def\eeq{\end{equation}}
\def\bea{\begin{eqnarray}}
\def\eea{\end{eqnarray}}

\title{Production of two $c \bar c$ pairs in gluon-gluon scattering 
\\ in high energy proton-proton collisions}

\author{W. Sch\"afer}%
\affiliation{Institute of Nuclear Physics, PL-31-342 Cracow, Poland}

\author{A. Szczurek}
\affiliation{Institute of Nuclear Physics, PL-31-342 Cracow, Poland}
\affiliation{Rzesz\'ow University, PL-35-959 Rzesz\'ow, Poland}

\date{\today}%

\begin{abstract}
We calculate cross sections for $g g \to Q \bar Q Q \bar Q$ in the high-energy
approximation in the mixed (longitudinal momentum fraction, impact parameter)
and momentum space representations.
Besides the  total cross section as a function of subsystem energy 
also differential distributions (in quark rapidity, transverse momentum, 
$Q Q$, $Q \bar Q$ invariant mass) are presented.
The elementary cross section is used to calculate production of 
$(c \bar c) (c \bar c)$ in single-parton scattering (SPS) in 
proton-proton collisions. 
We present integrated cross section as a function of proton-proton
center of mass energy as well as differential distribution
in $M_{(c \bar c)(c \bar c)}$. The results are compared with
corresponding results for double-parton scattering (DPS) discussed
recently in the literature.
We find that the considered SPS contribution to $(c \bar c)(c \bar c)$
production is at high energy ($\sqrt{s} >$ 5 TeV) much smaller than that
for DPS contribution.
\end{abstract}

\maketitle

\section{Introduction}

The cross section for $c \bar c$ production in proton-proton
or proton-antiproton collisions at high energy is quite large
\cite{book,LMSccbar2011} because the gluon-gluon luminosity grows quickly
with energy. 
We have shown recently that the cross section
for production of two pairs of $c \bar c$ in double-parton scattering
(DPS) grows even faster with incident center-of-mass energy \cite{LMSDPS2011}
and becomes very large at large energies.
In order to verify the DPS contribution a single-parton scattering (SPS)
contribution has to be evaluated. This was not done so far in the
literature as it requires calculation of $2 \to 4$ subprocesses.
At LHC a measurement of the two-pairs of $c \bar c$ production should 
be possible. This could be a good test of methods of calculating
higher-order QCD corrections.
 
It is the aim of the present paper
to calculate contribution of single-parton scattering to
the inclusive $p p \to (c \bar c) (c \bar c) X$ cross section.
In the present paper we shall use high-energy approximation in calculating
elementary $g g \to (c \bar c)(c \bar c)$ cross section.
At low incident energy and/or low $c \bar c c \bar c$ invariant mass
production a carefull treatment of the threshold effects is required.
The elementary $2 \to 4$ cross section is convoluted next with
gluon distribution functions. The result is compared with that
for DPS presented recently \cite{LMSDPS2011}. A prospects how to
disentangle SPS and DPS contributions will be discussed in the Result
section.

\section{Theoretical framework}

\begin{figure}[h!]
\includegraphics[width=6cm]{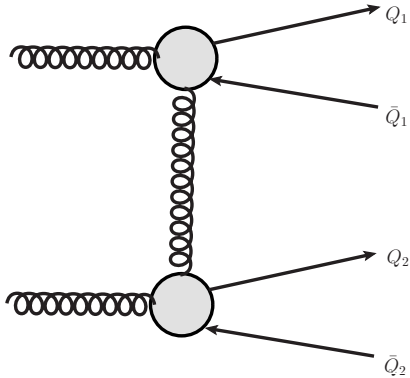}
\caption{
One-gluon exchange amplitude for the $g g \to (c \bar c)(c \bar c)$ process.}
\label{fig:gg_QQbarQQbar}
\end{figure}

\subsection{$gg \to Q\bar Q Q \bar Q$}

In the high energy limit of the gluon-gluon subprocess, the $gg \to Q\bar Q Q \bar Q$
amplitude will be dominated by the $t$-channel gluon exchange. 
Furthermore, if  the $Q \bar Q$-pairs are produced in the respective 
fragmentation regions of the incoming gluons an intuitively appealing approach based
on light-cone perturbation theory is possible. Namely we are looking for the cross section of 
excitation of the $Q \bar Q$-Fock states of the colliding gluons.

We base our calculations on previous works \cite{NPZ94,NPZ96}, where the total cross section 
for the process $g N \to Q \bar Q X$ and $g A \to Q \bar Q X$  has been obtained, 
as well as  on \cite{SingleJet}, where more detailed kinematical distributions can be found.

Let us briefly recapitulate the derivation of the key formulas.
We start by writing out the lightcone Fock-state expansion of the incoming, physical,  gluon, 
taking into account only the heavy-quark-antiquark fluctuation in the two-body sector
(see, \cite{NPZ96}):

\begin{eqnarray}
\ket{g^a(\bb)} = \sqrt{1-n_{Q \bar Q}} \, \ket{g^a_{\mathrm{bare}}} + \int d^2\br dz \, \Psi(z,\br)
\ket{[Q \bar Q]_8^a;z,\br} \; .
\label{eq:Fock}
\end{eqnarray}
Here, quark and antiquark in the gluon carry fractions $z,1-z$ of the gluon's 
large light-cone plus-momentum and are seperated by a distance $\br$ in the impact 
parameter plane. Notice that the quark-antiquark pair before the interaction 
is \emph{not} a color dipole, but carries a color-octet charge. 
Infrared safety of the relevant cross section is a consequence of the fact, that 
arbitrarily long-wavelength gluons cannot induce the $g \to Q\bar Q$ 
transition and decouple.

The normalized color-states of the quark-antiquark system in the color-octet and 
color-singlet states are, respectively:
\begin{eqnarray}
\ket{[Q \bar Q]^a_8} = \sqrt{2} \, (t^a)^i_j \, \ket{ Q_i \bar Q^j} 
\, , \, 
\ket{[Q \bar Q]_1} = {1 \over \sqrt{N_c}} \, \delta^i_j \,  \ket{Q_i \bar Q^j} \, .
\end{eqnarray}
A similar Fock-space expansion as (\ref{eq:Fock}) can be written for the second   
gluon participating in the interaction, which will have a large momentum
in the light-cone minus-direction. 

The interaction between the right- and
left moving parton systems is then mediated by the gluon exchange which acts like
a helicity conserving potential \cite{Gunion_Soper} between partons $i$ and $j$.

\begin{eqnarray}
V(\bb + \bb_i - \bs_j) = (-i) {\alpha_S \over \pi} \, \int {d^2 \bq \over
[\bq^2 + \mu_G^2]} \, \exp[i(\bb+\bb_i - \bs_j)\bq] \, T^b_i \otimes T^b_j \, .
\end{eqnarray}

\begin{figure}[h!]
\includegraphics[width=12cm]{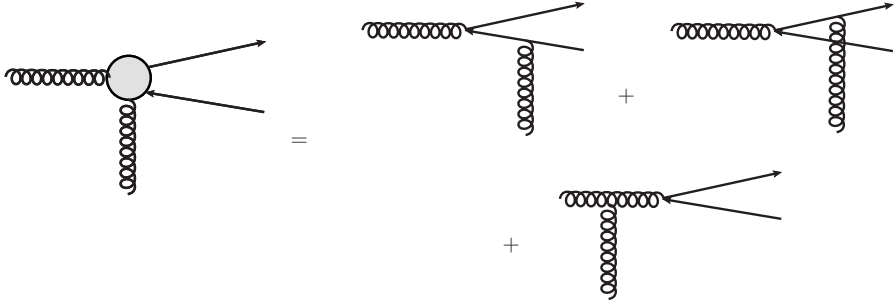}
\caption{
Feynman diagrams for the $g \to Q {\bar Q}$ impact factor.}
\label{fig:impact_factor}
\end{figure}

Here $\bb$ is the impact parameter between the two initial gluons, $\bb_i$
and $\bs_i$ are the impact parameter distances of $Q$ and $\bar Q$ 
relative to their respective parent gluon which are conserved 
during the interaction. The matrices $T^b_i$ are the generators of color $SU(N_c)$
acting on parton $i$ in the relevant representation.
Notice, that the gluon mass parameter $\mu_G$ is not needed to make the cross section 
finite, as mentioned above there are no singularities associated with the small-$\bq$ 
behaviour of the gluon propagator. As a physical parameter it enforces a finite
propagation radius of gluons in the transverse plane, as is in fact 
enforced by confinement. In practice, for the problem at hand its precise value
is unimportant: as long as $\mu_G^2 \ll m_Q^2$ our results are practically independent
of $\mu_G$.
The relevant Feynman diagrams for our process are shown in 
Figs.(\ref{fig:gg_QQbarQQbar}) and (\ref{fig:impact_factor}),
and the corresponding scattering amplitude for $gg \to (Q \bar Q)(Q \bar Q)$ 
in the high-energy limit of interest takes the form:
\begin{eqnarray}
[S(\bb) -1]\ket{g^a \otimes g^c;\bb} = 
-i {\alpha_S \over \pi}
\int { d^2 \bq e^{i \bq \bb} \over [\bq^2 + \mu_G^2]} \,
\Big[\sum_j \,  e^{-i \bq \bs_j} \, T^b_j \ket{g^c} \Big]  
\otimes \Big[\sum_k \,  e^{i \bq \bb_k} \, T^b_k \ket{g^a} \Big]  
\; . \nonumber \\
\end{eqnarray}
Let us now turn to one of the factors in the square brackets (the
so-called ``impact-factors''). The quark in the gluon is located 
at the distance $\bb_Q = (1-z) \br$, and
the antiquark at $\bb_{\bar Q} = - z \br$, where $\br = \bb_Q - \bb_{\bar Q}$.
In fact we are interested in the respective two-body Fock-state components 
orthogonal to the physical gluon, and
the relevant piece of the amplitude for the excitation $g \to Q \bar Q$ 
is then given by (see also \cite{NPZ96}):

\begin{eqnarray}
\Phi^b(\bq, g^a) &&\equiv 
\sum_k \,  e^{i \bq \bb_k} \, T^b_k \ket{G^a} \Big|_{\mathrm{excitation}}
\nonumber \\
&&= \int dz d^2\br \Psi(z,\br) 
\Big\{
[ e^{i (1-z) \br\bq} - e^{-i z\br\bq} ] \, {1 \over {\sqrt{2 N_c}}} \, \delta_{ab} 
\ket{[Q\bar Q]_1;z,\br}
\nonumber \\
&&+ [ e^{i (1-z) \br\bq} - e^{-i z\br\bq} ] \, {1 \over 2} \, 
d_{abc}  \ket{[Q \bar Q]^c_8;z,\br} 
\nonumber \\
&& +  [ e^{i (1-z) \br\bq} + e^{-i z\br\bq} -2 ] 
 {1 \over 2} \, 
i f_{abc}  \ket{[Q \bar Q]^c_8;z,\br} 
\Big\}  \; .
\end{eqnarray}
Here $b$ is the color index of the $t$-channel gluon which carries the transverse
momentum $\bq$. Clearly, the impact factor vanishes for $\bq \to 0$.
A completely analogous expression can be written for the other factor
in square brackets, and the full amplitude is obtained as:
\begin{eqnarray}
A(g^a g^c \to Q \bar Q Q \bar Q;\bb) = -i {\alpha_S \over \pi}
\int {d^2\bq e^{i \bq \bb} \over [\bq^2 + \mu_G^2]} 
\Phi^b(\bq,g^a) \, \Phi^b(-\bq,g^c) \, .
\end{eqnarray}
The two impact factors correspond to the upper and lower gluons, respectively.
The total cross section is then obtained after integrating
the squared amplitude over the impact parameter and averaging over initial
gluon colors
\begin{eqnarray}
&&\sigma_{tot} = {1 \over (N_c^2 -1)^2} \sum_{a,c} \int d^2 \bb 
|A(g^a g^c \to Q \bar Q Q \bar Q;\bb)|^2 =
4 \, \alpha_S^2 
\int {d^2\bq I^{bc}(\bq) \, I^{bc}(-\bq) \over [\bq^2 + \mu_G^2]^2} 
\, .
\end{eqnarray}
Similar impact factor representations for related QED problems are known for a long
time \cite{Cheng_Wu}.
Introducing the short-hand notation
\begin{eqnarray}
F(\bq \br) = 2 - \exp(i\bq \br) - \exp(-i \bq \br) \, ,
\end{eqnarray}
after some calculation, we obtain the impact factors relevant for the total 
cross section:
\begin{eqnarray}
I^{bc}(\bq) &&\equiv {1 \over N_c^2 -1} 
\sum_a \Phi^b(\bq, g^a) \Phi^{c*}(\bq,g^a)
\nonumber \\
&&= 
{\delta_{bc} \over 2 N_c}{N_c^2 \over  N_c^2 -1} \int dz d^2\br |\Psi(z,\br)|^2 
\Big( 
F((1-z)\bq\br) + F(z\bq\br) - {1 \over N_c^2} F(\bq\br)
\Big) \, .
\end{eqnarray}
The light-cone wave function for the $g \to Q \bar Q$ transition can be obtained from the
well-known case for the photon as \cite{NZ90,NPZ96}:
\begin{equation}
|\Psi(z,\br)|^2 = {\alpha_S(r) \over 6 \alpha_{em} } |\Psi_{\gamma}(z, \br) |^2
= {\alpha_S(r) \over (2 \pi)^2} \, 
\Big[ \Big(z^2+(1-z)^2 \Big) m_Q^2 K_1^2(m_Q r) + m_Q^2 K_0^2(m_Q r) \Big] \, , 
\end{equation}
where $K_{0,1}$ are generalized Bessel functions, and in the spirit of collinear
factorization, we took the gluon to be on-shell.

\subsection{Dipole-dipole cross section}

It is now convenient to introduce the total cross section for two color dipoles
of sizes $\br$, $\bs$, in the two-gluon exchange Born-approximation \cite{NZZ93}:

\begin{eqnarray}
\sigma_{DD}(\br,\bs) &&= {N_c^2-1 \over N_c^2} \, \int \, {d^2\bq \, \alpha_S^2 \over [\bq^2 + \mu_G^2]^2}
\, F(\bq \br) F(-\bq \bs) 
\nonumber \\
&&=  4  {N_c^2-1 \over N_c^2} \alpha_S^2 
\Big( \chi(0) - \chi(\br) -\chi(\bs) + \chi(\br - \bs) \Big) \, ,
\end{eqnarray}
where
\begin{eqnarray}
\chi(\br) &&=  \int \, {d^2\bq \over [\bq^2 + \mu_G^2]^2} \, \exp(i \bq\br) 
= {\pi \over \mu_G^2} \, (\mu_G r) \, K_1(\mu_G r) \, .
\end{eqnarray}
The Born level dipole-dipole cross section now reads
\begin{eqnarray}
\sigma_{DD}(\br,\bs) &&= {N_c^2-1 \over N_c^2} \, {4 \pi \alpha_S^2  \over \mu_G^2} \,
\Big[ 1 - \mu_G r K_1(\mu_G r) - \mu_G s K_1(\mu_G s) + \mu_G|\br - \bs| K_1( \mu_G|\br - \bs|)
\Big] \, .  \nonumber \\
\end{eqnarray}
Notice, that it is finite for $\mu_G \to 0$.

The total cross section for the parton-level process $gg \to Q \bar Q Q \bar Q$ can now
be written in terms of the dipole-dipole cross section and the light-cone wave-functions for the
$g \to Q \bar Q$ transitions as
\begin{eqnarray}
&&\sigma_{tot} = \int dz d^2\br du d^2\bs |\Psi(z,\br)|^2  |\Psi(u,\bs)|^2
\, \Sigma(z,\br;u,\bs) \, .
\label{sigma_dipole_dipole}
\end{eqnarray}
where
\begin{eqnarray}
\Sigma(z,\br;u,\bs) \, .
&&= 
\Big({N_c^2 \over N_c^2-1}\Big)^2 
\nonumber \\
&&\cdot 
\Big\{
\sigma_{DD}((1-z)\br,(1-u)\bs)
+\sigma_{DD}((1-z)\br,u\bs)
-{1 \over N_c^2} \sigma_{DD}((1-z)\br,\bs)
\nonumber \\
&&+ 
\sigma_{DD}(z\br,(1-u)\bs)
+\sigma_{DD}(z\br,u\bs)
-{1 \over N_c^2} \sigma_{DD}(z\br,\bs)
\nonumber \\
&&-
{1 \over N_c^2}\Big(
\sigma_{DD}(\br,(1-u)\bs)
+\sigma_{DD}(\br,u\bs)
-{1 \over N_c^2} \sigma_{DD}(\br,\bs)
\Big)
\Big\} \, .
\end{eqnarray}

\subsection{Momentum distributions}

The mixed representation given above could in principle be used to
obtain distributions in longitudinal momenta, simply by stripping
off the $dz$ and/or $du$ integrations. For the more interesting
transverse momentum distributions it is better to start all over
in momentum space, where the impact factor reads
\begin{eqnarray}
\Phi^b(\bq,g^a) &&= \int dz {d^2\bk \over (2 \pi)^2}
\Big\{
[ \Psi(z,\bk - (1-z) \bq) - \Psi(z,\bk + z \bq) ] \, {1 \over {\sqrt{2 N_c}}} \, \delta_{ab} 
\ket{[Q\bar Q]_1;z,\bk}
\nonumber \\
&&+
[ \Psi(z,\bk - (1-z) \bq) - \Psi(z,\bk + z \bq) ] 
\, {1 \over 2} \, 
d_{abc}  \ket{[Q \bar Q]^c_8;z,\bk} 
\nonumber \\
&&+ 
[ \Psi(z,\bk - (1-z) \bq) - \Psi(z,\bk + z \bq) -2 \Psi(z,\bk)] 
 {1 \over 2} \, 
i f_{abc}  \ket{[Q \bar Q]^c_8;z,\bk} 
\Big\} \; .
\end{eqnarray}
After squaring, we obtain naturally the same structure as in \cite{SingleJet}:
\begin{eqnarray}
I^{bc}(z,\bk,\bq) 
&&= {\delta_{bc} \over 2 N_c} \Big\{ 
{N_c^2 \over N_c^2 -1} \Big[
| \Psi(z,\bk) - \Psi(z,\bk + z \bq) |^2
+ | \Psi(z,\bk) - \Psi(z,\bk -(1- z) \bq) |^2
\nonumber \\
&&
- | \Psi(z,\bk-(1-z)\bq) - \Psi(z,\bk +z \bq) |^2
\Big]
+  | \Psi(z,\bk-(1-z)\bq) - \Psi(z,\bk +z \bq) |^2
\Big \} \, .
\nonumber \\
&&\equiv {\delta_{bc} \over 2 N_c} \, I(z,\bk,\bq)
\end{eqnarray}
The explicit form for the squares of the light-cone wave functions can
been found in \cite{SingleJet} and reads:
\begin{eqnarray}
\Big| \Psi(z,\bp) - \Psi(z,\bp+ \bkappa) \Big|^2 &&= \alpha_S \, \Big\{
[ z^2 + (1-z)^2 ] \Big( {\bp \over \bp^2 + m_Q^2}  - {\bp + \bkappa 
\over (\bp + \bkappa)^2 + m_Q^2} \Big)^2 
\nonumber \\
&&+ m_Q^2 \Big( {1 \over \bp^2 + m_Q^2}  -
{ 1 \over (\bp + \bkappa)^2 + m_Q^2} \Big)^2 
\Big\} \; .
\label{part_of_impact_factor}
\end{eqnarray}
With all of this given, we can put together 
the differential cross section
\begin{eqnarray}
d \sigma &&=  
{N_c^2-1 \over N_c^2} \, 
{4\pi^2 \alpha_S^2 \over [\bq^2 + \mu_G^2]^2} \,
 I(z,\bk,\bq) I(u,\bl,-\bq) \, 
dz {d^2\bk \over (2 \pi)^2} \, du   {d^2\bl \over (2 \pi)^2} 
{d^2 \bq \over (2 \pi)^2}
\, . \nonumber \\
\label{sigma_momentum_space}
\end{eqnarray}
A brief comment on the kinematics of quarks is in order. 
The four-momenta of quarks are fully specified by the variables
$z,\bk,\bq,u,\bl$.
Incoming gluons carry longitudinal momentum fractions $x_1,x_2$ of the momenta 
of the incoming protons. The latter are
\begin{eqnarray}
P_a = P_a^+ n^+ = \sqrt{{s \over 2}} n^+, \, \, \, P_b = P_b^- n^- = 
\sqrt{{s \over 2}}  n^- \, ,
\end{eqnarray}
so that $2 P_a \cdot P_b = s$.
We parametrize four-momenta in light-cone coordinates
\begin{eqnarray}
p = (p_+,p_-, \bp), \, \,  \, 2 p_+p_- - \bp^2 = m^2  \; .
\end{eqnarray}
Then, the four momentum, say of the quark/antiquark belonging to parent 
gluon $1$, read
\begin{eqnarray}
p_{Q_1} = (zx_1P_a^+, {\bp_Q^2 + m_Q^2 \over 2 z x_1 P_a^+}, \bp_Q), \, \, \, 
p_{\bar{Q}_1} = ((1-z)x_1P_a^+, {\bp_{\bar Q}^2 + m_Q^2 \over 2 (1-z) x_1 P_a^+}, \bp_{\bar Q}) \, .
\end{eqnarray}
We still need to give the relation between the transverse momenta 
$\bp_{Q, \bar Q}$ of the
(anti-)quark and the momenta $\bk,\bq$ used above.
These relations read:
\begin{eqnarray}
\bp_Q = \bk + z \bq, \, \, \, \bp_{\bar Q} = -\bk + ( 1-z) \bq \, ,
\end{eqnarray}
or, alternatively
\begin{eqnarray}
\bq = \bp_{Q} + \bp_{\bar Q} , \, \, \, \bk = (1-z) \bp_Q - z \bp_{\bar Q} \, .
\end{eqnarray}
Notice, that $\bq$ is the total transverse momentum of the $Q \bar
Q$-pair, while $\bk$ is the light-cone relative transverse momentum
which was conjugate to the dipole size.
Now having the four momenta, we can calculate all sorts
of kinematical variables, e.g. the rapidity will be given by
\begin{eqnarray}
y_Q = {1 \over 2} \log \Big( {p_Q^+ \over p_Q^-} \Big) \, . 
\end{eqnarray}

The kinematics of the ``lower'' $Q \bar Q$-pair is treated analogously. 
Obviously the transverse momentum of the second pair is just $-\bq$, because
incoming gluons are collinear.
When constructing four-momenta, we only need to be careful that the 
large component is along the lightcone-minus direction
\begin{eqnarray}
p_{Q_2} = (  {\bp_Q^2 + m_Q^2 \over 2 u x_2 P_b^-}
, ux_2 P_b^-, \bp_Q), \, \, \, 
p_{\bar Q_2} = (
 {\bp_{\bar Q}^2 + m_Q^2 \over 2 (1-u) x_2 P_b^-},  
(1-u)x_2P_a^+, \bp_{\bar Q}) \, .
\end{eqnarray}
And here 
\begin{eqnarray}
\bp_Q = \bl - u \bq, \, \, \, \bp_{\bar Q} = -\bl - ( 1-u) \bq \, .
\end{eqnarray}

\subsection{On normalization and phase-space}

It is important to remember, that the expressions we derived are valid
in a high energy limit, in which the invariant masses of $Q \bar Q$-pairs
are much smaller than the center-of-mass energy squared 
of the gluon-gluon collision: $M_{12}^2, M_{34}^2 \ll \hat{s}$.

In the practical calculation of the hadron-level cross section, we
integrate over all momentum fractions carried by initial state gluons,
and hence over the cms-energy of the gluon-gluon subprocess. 
In practice, we want that our cross section behaves smoothly also
at low energies. 

Firstly notice, that our states are normalized in such a way, that the
formulas are simple in the high energy limit. In particular,
the constrained two-body phase space is just $dz d^2\bk/(2 \pi)^2$.

Effectively, the four-body phase space in the high-energy limit is just 
\begin{eqnarray}
{d^2\bq \over (2 \pi)^2} \, dz {d^2\bk \over (2 \pi)^2} \,  du {d^2\bl
  \over (2 \pi)^2}  \, .
\label{elementary_cross_section_integration}
\end{eqnarray}
Obviously it doesn't vanish if $\hat{s}$ approaches the threshold.
To improve upon this, let us first rescale the amplitude, and introduce
the Feynman-amplitude ${\cal{M}}_F$ 
\begin{eqnarray}
{\cal{M}}_F &&\equiv  2P_+ \, \sqrt{z(1-z)(2\pi)} \, 2P_- \sqrt{u(1-u)(2 \pi)} \cdot A 
\nonumber \\
&&=  \sqrt{z(1-z) 2 (2\pi)}  \sqrt{u(1-u)2(2\pi)} \cdot \hat{s} \cdot A \, ,
\nonumber \\
\end{eqnarray}
which is normalized such, that the fully differential cross section takes
the form:
\begin{eqnarray}
d \sigma = {1 \over 2 \hat{s}} \, |{\cal{M}}_F|^2 \,  d \Phi(\hat{s};p_1,p_2,p_3,p_4) \, .
\end{eqnarray}
Here, the constrained n-body phase space, with $P_{in}^2 = \hat{s}$ is
\begin{eqnarray}
 d \Phi(\hat{s};p_1,\dots,p_n) = (2 \pi)^4 \delta^{(4)}(P_{in} - \sum_i p_i ) \cdot 
\prod_{i=1}^n {d^4 p_i \over (2 \pi)^3} \, \delta(p_i^2 - m_i^2) \, .
\end{eqnarray}

We now introduce invariant masses $M_{12}, M_{34}$, and write the four-body phase-space
as 
\begin{eqnarray}
 d \Phi(s;p_1,p_2,p_3,p_4) =  d \Phi(s;P_{12},P_{34}) \, {d M_{12}^2 \over 2 \pi} 
 d \Phi(M_{12}^2;p_1,p_2) \, {d M_{34}^2 \over 2 \pi}  d
 \Phi(M_{34}^2;p_3,p_4) \, .
\label{4body_PS}
\end{eqnarray}
Now, going over to light-cone coordinates, we have
\begin{eqnarray}
 {d M^2 \over 2 \pi} \,  d \Phi(M^2;p_1,p_2) &&=   {d M^2 \over 2 \pi} 
\, (2 \pi)^4 \delta^{(4)}(P_{12}- p_1 - p_2)  
\prod_{i=1}^2 {dp_i^+ dp_i^- d^2\bp_i \over (2 \pi)^3} \, \delta(2p_i^+ p_i^- - \bp_i^2 - m_i^2) 
\nonumber \\
&&=  {d M^2 \over 2 \pi} \, (2 \pi)^4 \, \delta(p_1^+ + p_2^+ - P_{12}^+) \, 
\delta({\bp_1^2 + m_1^2 \over 2 p_1^+}+ {\bp_2^2 + m_2^2 \over 2 p_2^+} - P_{12}^-)
\nonumber \\
&& \cdot \delta^{(2)}(\bp_1 + \bp_2 - \bP_{12}) \, \prod_{i=1}^2 {d
  p_i^+ d^2 \bp_i \over 2 p_i^+ (2 \pi)^3 } \; .
\end{eqnarray}
Here we integrated out the $p_i^-$-components from the on-shell conditions.
Let us now write $p_1^+ = z P_{12}^+$, then one of the overall delta-functions
gives us, that $p_2^+ = (1-z) P_{12}^+$. Furthermore, we can use the on-shell
condition $P_{12}^2 = M^2$ to write $2 P_{12}^+ P_{12}^- = M^2 + \bP_{12}^2$. then
we obtain finally
\begin{eqnarray}
 {d M^2 \over 2 \pi} \,  d \Phi(M^2;p_1,p_2) &&=  
{dz \over z (1-z) } \, {d^2 \bp_1 d^2 \bp_2 \over 2 (2 \pi)^3} \, 
 \delta^{(2)}(\bp_1 + \bp_2 - \bP_{12}) 
\nonumber \\
&&=  {dz \over z (1-z) } \, {d^2 \bk \over 2 (2 \pi)^3} \, .
\end{eqnarray}
Which agrees, modulo the factors now absorbed in the Feynman amplitude with 
the result obtained in the high-energy limit. 
The major simplification in the high-energy limit occurs in the first factor
of the four-body phase space (\ref{4body_PS}). Namely we can write 
the phase space for two clusters $M_{12}^2,M_{34}^2 \ll \hat{s}$. 
which are seperated by a large rapidity distance as
\begin{eqnarray}
 d \Phi(s;P_{1},P_{2})=  (2 \pi)^4 \, \delta^{(4)} (P - P_{1}-P_{2}) \, 
\prod_{i=1}^2 {dP_i^+ dP_i^- d^2\bP_i \over (2 \pi)^3} \, \delta(2P_i^+ P_i^- - \bP_i^2 - M_i^2) 
\, .
\end{eqnarray}
For brevity, we wrote $P_1 = P_{12}, P_2 = P_{34}$
Now we can neglect the plus-component of $P_2$ and 
the minus-component of $P_1$ in the overall four-momentum
conservation, so that
\begin{eqnarray}
 \delta^{(4)} (P - P_{1}-P_{2}) = \delta(P^+ - P_1^+) \delta(P^- - P_2^-) \delta^{(2)}(\bP_1 + \bP_2) \, .
\end{eqnarray}
Hence all the integrations except for the transverse
momentum ones can be done immediately. From the integrals over 
$\pm$-components, we only get a factor of 
\begin{eqnarray}
{1 \over 2 P_1^+ \, 2 P_2^-} = {1 \over 2 \hat{s}} \, ,
\end{eqnarray}
so that, finally:
\begin{eqnarray}
 d \Phi(\hat{s};P_{1},P_{2})= {1 \over 2 \hat{s}} \cdot {d^2 \bP_1 d^2 \bP_2 \over (2 \pi)^2} \,  
\delta^{(2)}(\bP_1 + \bP_2)  \equiv 
 {1 \over 2\hat{s}} \cdot {d^2 \bq \over (2 \pi)^2} \; . 
\end{eqnarray}
If we approach the threshold, the exact phase-space goes to zero and our 
approximation is very bad. 
Still, we know, that the \emph{integrated} two-cluster phase-space will be just
\begin{eqnarray}
\int  d \Phi(\hat{s};P_{1},P_{2})=
{\sqrt{[\hat{s}-(M_{12}+M_{34})^2][\hat{s}-(M_{12}-M_{34})^2]} \over 8
  \pi \hat{s}} \; .
\end{eqnarray}
In order to improve our calculation,
we therefore introduce the correction factor
\begin{eqnarray}
f_{\mathrm{corr}} = \sqrt{\Big[1-{(M_{12} + M_{34})^2 \over \hat{s}}\Big]
\Big[1-{(M_{12} - M_{34})^2 \over \hat{s}}\Big]} \, ,
\label{threshold_correction_factor}
\end{eqnarray}
which within the approximations of the high-energy limit of course is exactly unity.
A deviation from unity, or as a matter of fact any energy-dependence of the subprocess
cross section thus indicates for us how far we are from the high-energy domain.

\subsection{$p p \to (Q \bar Q) (Q \bar Q)$ inclusive cross section}

The cross section for proton-proton (proton-antiproton) 
can be calculated as usually in the parton model as
\begin{equation}
\sigma_{p p \to (Q \bar Q)(Q \bar Q)}(W) = 
\int d x_1 d x_2 \; g(x_1,\mu_F^2) \; g(x_2,\mu_F^2) \;
 \sigma_{gg \to (Q \bar Q) (Q \bar Q)}({\hat s}^{1/2}) \; ,
\label{parton_model}
\end{equation}
where $\hat{s} = x_1 x_2 W^2$.
The last factor is the elementary cross section discussed in the previous
sections. In calculating (\ref{parton_model}) we take into account kinematical
constraints and the threshold correction factor 
(\ref{threshold_correction_factor}). The elementary cross section is 
calculated first on a subsystem energy grid and a simple interpolation
is done then when using it in formula (\ref{parton_model}). We shall 
use different parton (gluon) distributions from the literature. 
The factorization scale of the gluon distribution in principle
depends on the kinematics of the final state quarks.
We use $\mu_F^2 = 4 m_Q^2$ when calculating the integral 
(\ref{parton_model}). 
The parton formula (\ref{parton_model}) is very useful to 
make differential distribution in invariant mass of 
the $(Q \bar Q) (Q \bar Q)$ system. 
In principle it will be desirable to obtain results for
open-charm mesons, but the inclusion of hadronization 
goes beyond the scope of the present paper where we wish 
to present only a first estimation of the cross section 
for SPS production of $(c \bar c) (c \bar c)$. 

\section{Results}

\subsection{$g g \to (c \bar c) (c \bar c)$}

To make final calculations we have to fix $\alpha_s$ in formulae
(\ref{part_of_impact_factor}) and (\ref{sigma_momentum_space}). We shall use
leading-order running strong coupling constant $\alpha_s(k_i^2+m_c^2)$ 
in impact factors (see Eq.(\ref{part_of_impact_factor})) and 
$\alpha_s(\mu_i^2)$ with $\mu^2 = \max(k_{it}^2+m_Q^2,q^2)$
for gluon exchange (see Eq.(\ref{sigma_momentum_space})).

\begin{figure}[!h]
\includegraphics[width=8cm]{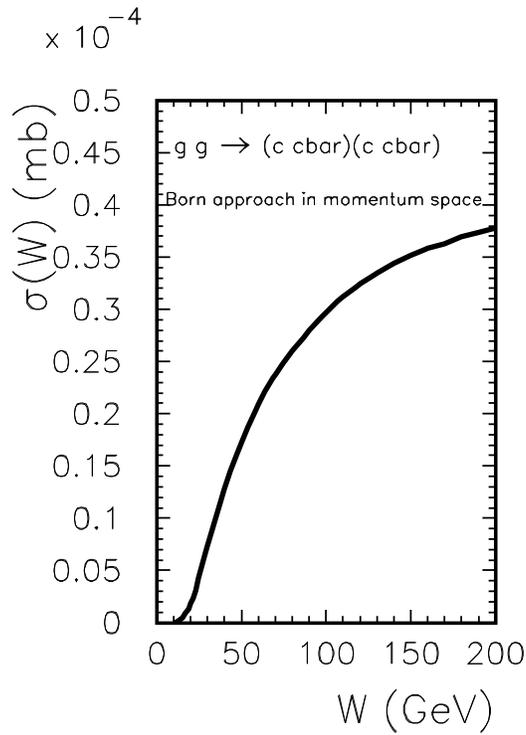}
\caption{Energy dependence of the total cross section
for the $g g \to (c \bar c) (c \bar c)$ process. The description of the
solid line is given in the main text.
}
\label{fig:sigma_gg_ccbarccbar_W}
\end{figure}

In Fig.\ref{fig:sigma_gg_ccbarccbar_W} we show total cross section
for the $g g \to (c \bar c)(c \bar c)$ as a function of gluon-gluon
energy. We show cross section with
extra cut $M_{12}+M_{34} < W$ 
and with extra correction factor 
(see Eq.(\ref{threshold_correction_factor})).
The latter cross section will be used then to calculate corresponding 
cross section for proton-proton collisions.
In the following calculations we have fixed the regularization parameter
$\mu_G$ = 0.5 GeV. There is only a marginal dependence on the value 
of the nonperturbative parameter.

\begin{figure}[!h]
\includegraphics[width=6cm]{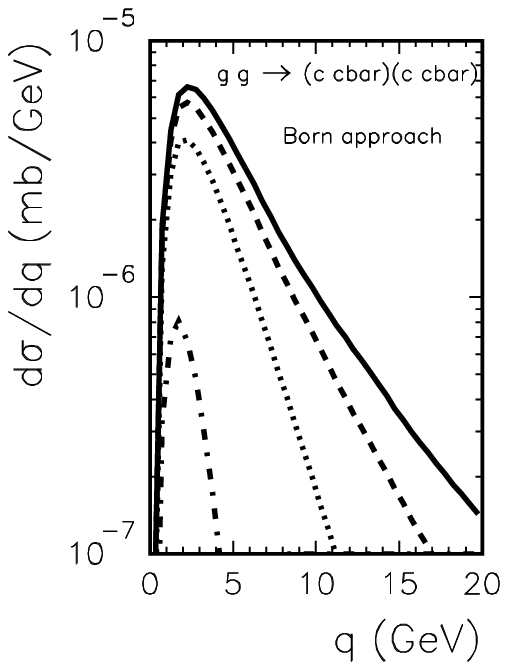}
\includegraphics[width=6cm]{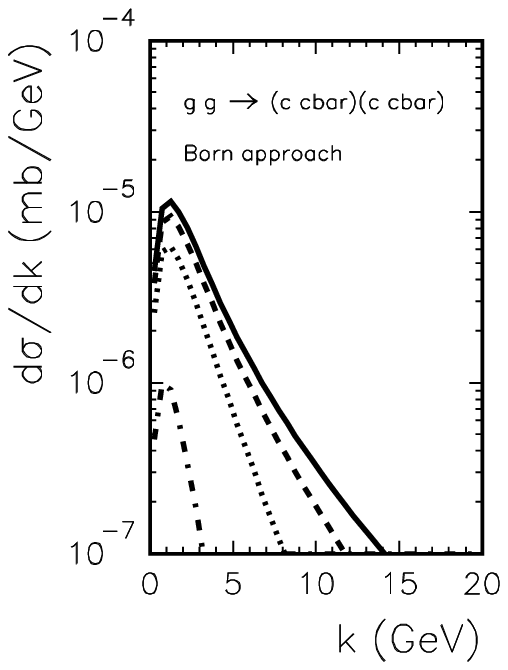}
\caption{Auxiliary distributions
for the $g g \to (c \bar c) (c \bar c)$ process
for $W$ = 20, 50, 100, 200 GeV (succesively growing cross section).
}
\label{fig:dsig_dq_and_dsig_dk}
\end{figure}

Before we go to real observables let us show an auxiliary distributions
in $q_t$ and $k_t$ (see Fig.(\ref{fig:dsig_dq_and_dsig_dk})).

\begin{figure}[h!]
\includegraphics[width=8cm]{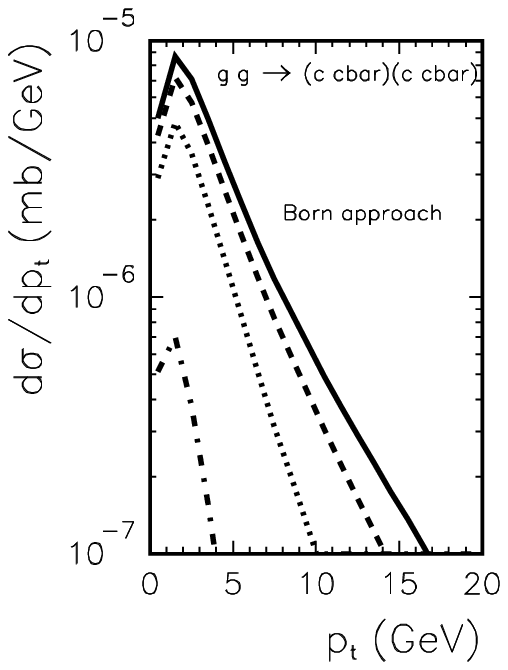}
\caption{The quark (antiquark) transverse momentum distribution
$d \sigma/dp_t$ in the $g g \to (c \bar c) (c \bar c)$ for 
$W =$ 20, 50, 100, 200 GeV.}
\label{fig:dsigma_dpt}
\end{figure}

The transverse momentum distributions of quark (antiquark) is shown
in Fig.\ref{fig:dsigma_dpt} for a few selected subsystem energies.
The higher subsystem energy the bigger transverse momenta are available
kinematically.

\begin{figure}[h!]
\includegraphics[width=8cm]{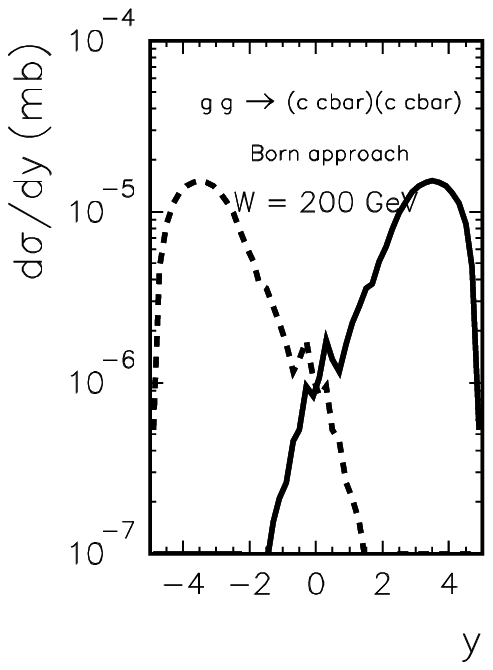}
\caption{Rapidity distribution of quark (antiquark) in the 
$g g \to (c \bar c) (c \bar c)$ for $W =$ 200 GeV.
The solid line is for emissions from the upper pair, the dashed line
is for emissions from the lower pair.} 
\label{fig:dsigma_dy}
\end{figure}

Rapidity distributions of quark (antiquark) for different energies are
shown in Fig.\ref{fig:dsigma_dy}. We show distributions for quarks emitted
from the upper line (solid line) and from the lower line (dashed line).
The higher the energy the better separation of the two contributions
can be seen.

\begin{figure}[h!]
\includegraphics[width=8cm]{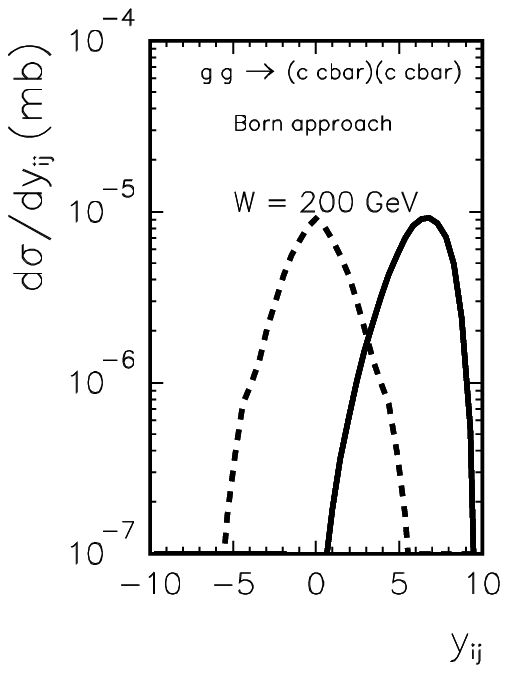}
\caption{Distribution in rapidity difference between quark-antiquark
from the same pair (dashed line) and quark-quark, quark-antiquark,
antiquark-quark, antiquark-antiquark from different pairs in the 
$g g \to (c \bar c) (c \bar c)$ for the subsystem energy $W =$ 200 GeV.} 
\label{fig:dsigma_dyij}
\end{figure}

The rapidity separation can be better seen in the distributions
in rapidity distance between (anti)quark-(anti)quark.
The distance between quark-antiquark from the same pair is very
small compared to the distance bewtween (anti)quark-(anti)quark
from different pairs. In Fig.\ref{fig:dsigma_dyij} we show
an example for subsystem energy $W$ = 200 GeV. The distance in rapidity
between quarks and antiquarks emitted from different pairs reminds a bit
situation in double-parton scattering. There the distance between quarks
and antiquarks emitted from two different hard processes can also be 
large \cite{LMSDPS2011}.

Finally we close presentation of our results for
$g g \to (c \bar c) (c \bar c)$ by showing distributions 
in quark-antiquark invariant masses.
Here there are two distinct classes of subsystems. We shall introduce 
the notation:
$ \frac{d \sigma}{dM_{12}} = \frac{d \sigma}{dM_{34}} \equiv
\frac{d \sigma^{I}}{dM_{ij}}$ (category I) \\
for quark-aniquark emitted in the same pair and \\
$ \frac{d \sigma}{dM_{13}} = \frac{d \sigma}{dM_{14}} =
  \frac{d \sigma}{dM_{23}} = \frac{d \sigma}{dM_{24}} \equiv
\frac{d \sigma^{II}}{dM_{ij}}$ (category II) \\
for quark-antiquark emitted from different pairs.
One can see that the average invariant mass of quark-antiquark
from the same pair is smaller than the average invariant mass from the
different pairs. At large invariant masses the emission from different
pairs dominates over the emission from the same pair.
The situation reminds that for double-parton scattering
\cite{LMSDPS2011}.

\begin{figure}[h!]
\includegraphics[width=8cm]{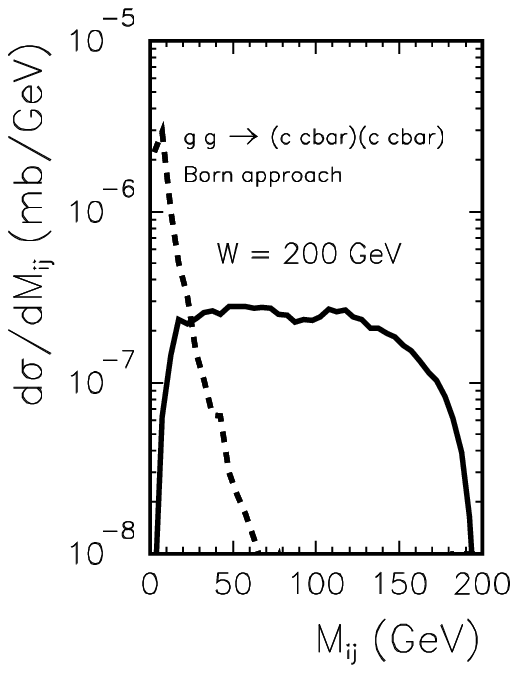}
\caption{Quark-antiquark invariant mass distributions
in the $g g \to (c \bar c)(c \bar c)$. The category I (same pair emission)
distributions are shown by the dashed line and category II (different
pair emission) distributions by the solid line.} 
\label{fig:dsigma_dM_ij}
\end{figure}

\subsection{$p p \to c \bar c c \bar c X$}

\begin{figure}[h!]
\includegraphics[width=8cm]{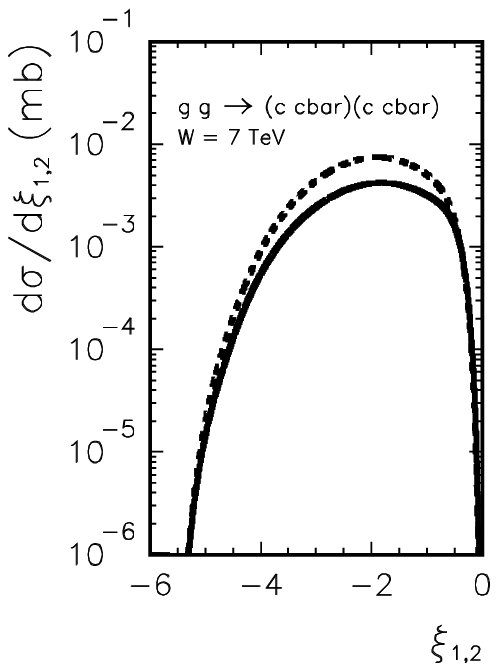}
\caption{Distribution of the proton-proton cross section in $\xi_1$
($\xi_2$) for the production of $(c \bar c)(c \bar c)$ at
$W=$ 7 TeV.
The solid line is for $\mu_F^2 = m_c^2$ and the dashed line is
for $\mu_F^2 = 4 m_c^2$. In this calculation a simple leading
order gluon distribution from \cite{CTEQ6} was used.
}
\label{fig:dsigma_dxi}
\end{figure}

Let us come now to proton-proton scattering.
In Fig.\ref{fig:dsigma_dxi} we show distribution in 
$\xi_1 = \log_{10}(x_1)$ ($\xi_2 = \log_{10}(x_2)$)
for $W=$ 7 TeV (LHC).
We see that typical $x_1$ and $x_2$ are not too small, of the order
of 10$^{-3}$ -- 10$^{-2}$. This is the region where the gluon
distributions are relatively well known.
There is no strong dependence of the cross section on the choice of
gluon distribution function (GDF).

\begin{figure}[h!]
\includegraphics[width=8cm]{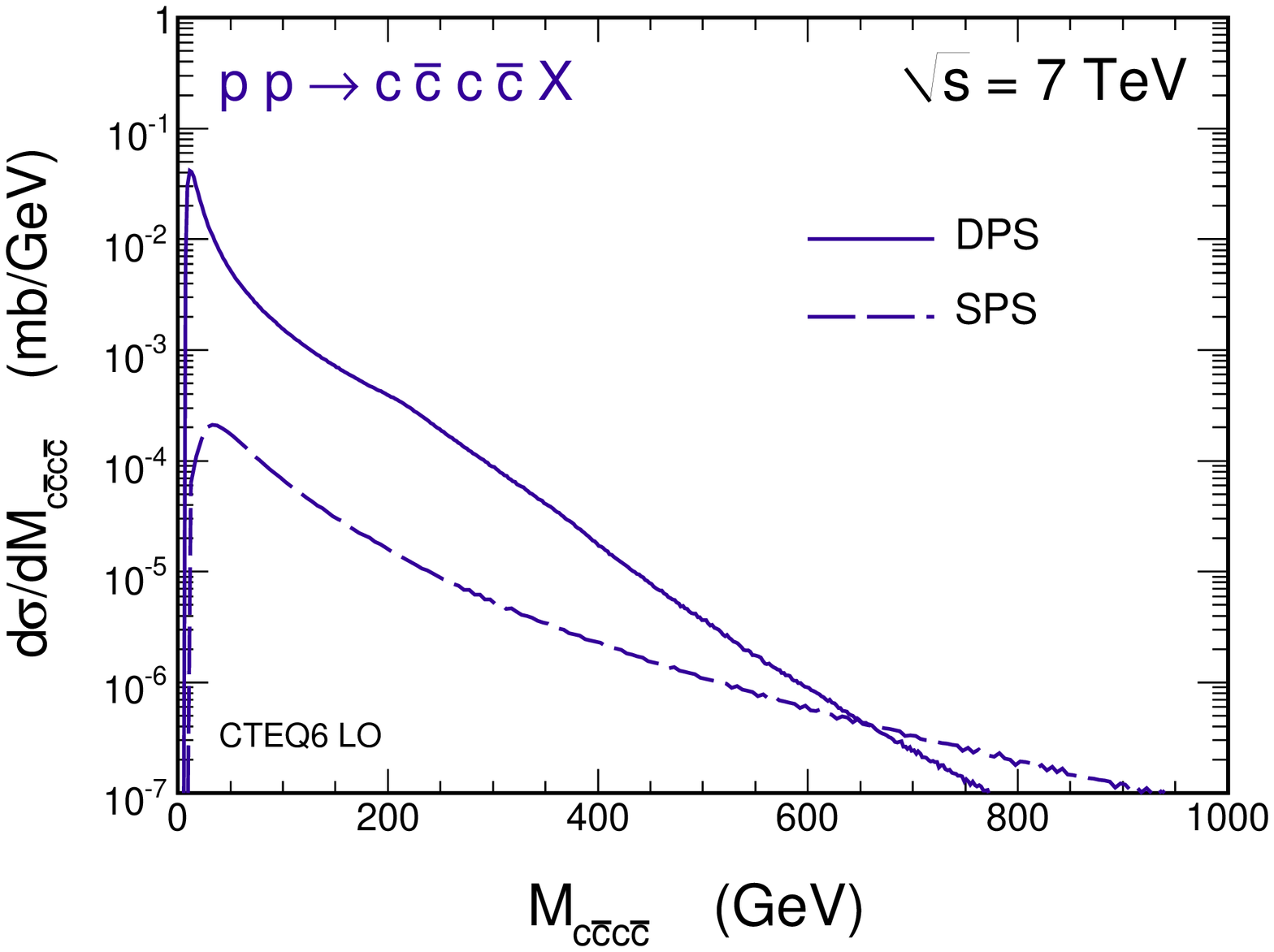}
\caption{
Distribution of the $p p \to c \bar c c \bar c X$ cross section 
in $M_{4c}$ for the production of $(c \bar c)(c \bar c)$ at
for $W=$ 7 TeV.
The solid line is for $\mu_F^2 = m_c^2$ and the dashed line is
for $\mu_F^2 = 4 m_c^2$. In this calculation a simple leading
order gluon distribution from \cite{CTEQ6} was used. 
}
\label{fig:dsigma_dM_4Q}
\end{figure}

The distribution in invariant mass of the $(c \bar c) (c \bar c)$
system (equal to subsystem energy) is shown in
Fig.\ref{fig:dsigma_dM_4Q}.
We show distribution for CTEQ6 GDFs \cite{CTEQ6}. For comparison we show
distribution obtained for double-parton scattering (see \cite{LMSDPS2011}).
While the double parton scattering contribution dominates in the region 
of small invariant masses, the single parton scattering contribution
takes over above $M_{4c} >$ 500 GeV. This is the region where large
rapidity gaps between quarks and antiquarks occur.


\begin{figure}[h!]
\includegraphics[width=8cm]{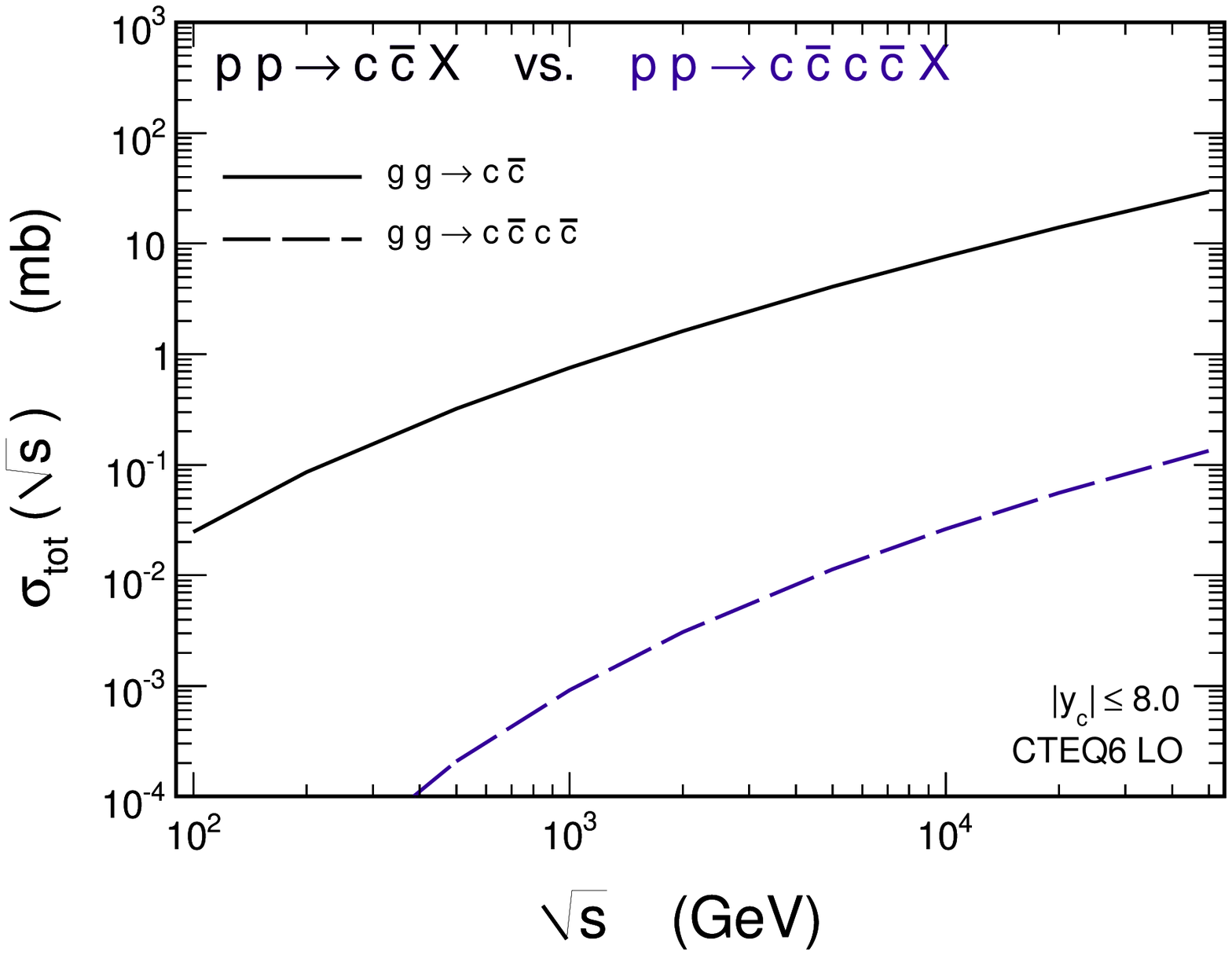}
\caption{Energy dependence of the cross section for 
the $p p \to c \bar c c \bar c X$ reaction
compared with that for the $p p \to c \bar c X$ reaction.
}
\label{fig:sigma_pp_ccbarccbar_W}
\end{figure}

The energy dependence of the $p p \to c \bar c c \bar c X$ inclusive
cross section is shown in Fig.\ref{fig:sigma_pp_ccbarccbar_W}.
One can observe that the inclusive cross section for the 
$c \bar c c \bar c$ final state is much smaller than that for 
the $c \bar c$ final state but grows somewhat faster at low energies. 
At higher energies the ratio is almost constant of the order of 1\%.
This is in contrast to double parton scattering contribution
for the $c \bar c c \bar c$ production which grows much faster
than the cross section for single $c \bar c$ production \cite{LMSDPS2011}. 

\section{Conclusions}

We have presented for the first time formulae for the production
of two pairs of $Q \bar Q$ in single-parton scattering.
The elementary $g g \to (Q \bar Q) (Q \bar Q)$ cross section was
given in two different representations: so-called mixed one 
(longitudinal momentum fraction, impact parameter) 
called also dipole representation, 
and momentum space one within a high (subsystem) energy approximation.
While the dipole representation is easy to include energy dependence
of the dipole-dipole interaction the momentum representation seems
better suited to include threshold effects. 
We have discussed how to correct the high-energy formulae
close to threshold where the phase space is rather limited by
energy-momentum conservation. 
 
We have presented energy dependence as well as different differential 
distributions for the elementary cross section for 
$g g \to (c \bar c) (c \bar c)$. We have shown that the elementary
cross section varies quickly from the kinematical threshold ($W = 4 m_c$)
up to $W =$ 100 GeV where almost a plateau can be observed.
The invariant mass distributions
$\frac{d \sigma}{dM_{12}} = \frac{d \sigma}{dM_{34}}$ are much
steeper than that for
$\frac{d \sigma}{dM_{13}} = \frac{d \sigma}{dM_{14}} =
 \frac{d \sigma}{dM_{23}} = \frac{d \sigma}{dM_{24}}$.
where 1 and 2 are from the first $c \bar c$ pair and 3 and 4 are from
the other $c \bar c$ pair.

Our high-energy approach does not include processes when a second pair
is produced via gluon emitted from quark/antiquark of a first pair
and its ``subsequent'' splitting into $c \bar c$. This processes
may be important for small rapidity distance between quarks and
antiquarks. A good example can be LHCb kinematics when both
$c$ quarks (or both $\bar c$ antiquarks) are emitted within
two units of rapidity. This mechanism should not be, however, important 
for the case of large rapidity interval between $c c$ or 
$\bar c \bar c$ discussed recently in the context of double parton
scattering \cite{LMSDPS2011}.

The elementary cross section has been convoluted next with gluon 
distributions in the proton. 
We have presented inclusive cross section for 
$p p \to (c \bar c) (c \bar c) X$ as a function of incident energy 
as well as invariant mass distribution of the $c \bar c c \bar c$
system. The results have been compared with corresponding
contribution of double-parton scattering.
We have found that the single-parton scattering contribution is 
significantly smaller than that for the double-parton scattering.
We conclude therefore that a measurement of two pairs of $c \bar c$
at LHC would be very useful in testing models of double parton scattering.

An evalution of distributions for charmed mesons would be very useful
in planning and interpreting current measurements at the LHC.
The LHCb collaboration started already such an analysis \cite{Belaev2012}.

\vspace{1cm}

{\bf Acknowledgment}

We are indebted to Rafa{\l} Maciu{\l}a for a discussion of double parton
scattering contribution and help in preparation of some figures 
and Vanja Belaev for the discussion of LHCb measurement. This work was 
partially supported by the polish MNiSW grant DEC-2011/01/B/ST2/04535.


\end{document}